\documentclass[a4paper,11pt]{article}

\usepackage{jheppub} 

\usepackage[T1]{fontenc} 

\usepackage{multirow}
\usepackage{subfigure}
\usepackage{float}
\usepackage{rotating}
\usepackage{color}

\newcommand{\pip}{\pi^+}
\newcommand{\pim}{\pi^-}
\newcommand{\piz}{\pi^0}
\newcommand{\etap}{\eta^{\prime}}

\newcommand{\psip}{\psi(2S)}

\newcommand{\jpsi}{J/\psi}

\newcommand{\EE}{e^+e^-}
\newcommand{\ee}{e^+e^-}
\newcommand{\MM}{\mu^+\mu^-}
\newcommand{\LL}{\ell^+\ell^-}
\newcommand{\pp}{\pi^+\pi^-}

\newcommand{\ppjpsi}{\pi^+\pi^- \jpsi}
\newcommand{\pppsip}{\pi^+\pi^- \psip}
\newcommand{\pphc}{\pp h_{c}}

\newcommand{\beq}{\begin{equation}}
\newcommand{\eeq}{\end{equation}}
\newcommand{\bitm}{\begin{itemize}}
\newcommand{\eitm}{\end{itemize}}

\newcommand{\gev}{\mathrm{GeV}}

\newcommand{\gevcc}{\mathrm{GeV}/c^2}
\newcommand{\dz}{D^{0}}
\newcommand{\dsm}{D^{*-}}

\title{\boldmath Observation of $\EE\to \eta\psip$ at center-of-mass energies from 4.236 to 4.600~GeV}

\collaboration{The BESIII Collaboration}
 \author{
M.~Ablikim$^{1}$, M.~N.~Achasov$^{10,c}$, P.~Adlarson$^{67}$, S. ~Ahmed$^{15}$, M.~Albrecht$^{4}$, R.~Aliberti$^{28}$, A.~Amoroso$^{66A,66C}$, M.~R.~An$^{32}$, Q.~An$^{63,49}$, X.~H.~Bai$^{57}$
% ...

\vspace{0.2cm} {\it
$^{1}$ Institute of High Energy Physics, Beijing 100049, People's Republic of China
 
% ...
$^{4}$ Bochum Ruhr-University, D-44780 Bochum, Germany
 
% ...
$^{10}$ G.I. Budker Institute of Nuclear Physics SB RAS (BINP), Novosibirsk 630090, Russia
 
% ...
$^{15}$ Helmholtz Institute Mainz, Staudinger Weg 18, D-55099 Mainz, Germany
 
% ...
$^{28}$ Johannes Gutenberg University of Mainz, Johann-Joachim-Becher-Weg 45, D-55099 Mainz, Germany
 
% ...
$^{32}$ Liaoning Normal University, Dalian 116029, People's Republic of China
 
% ...
$^{49}$ State Key Laboratory of Particle Detection and Electronics, Beijing 100049, Hefei 230026, People's Republic of China

% ...
$^{57}$ University of Jinan, Jinan 250022, People's Republic of China

% ...
$^{63}$ University of Science and Technology of China, Hefei 230026, People's Republic of China

% ...
$^{66}$ University of Turin and INFN, (A)University of Turin, I-10125, Turin, Italy; (B)University of Eastern Piedmont, I-15121, Alessandria, Italy; (C)INFN, I-10125, Turin, Italy\\
$^{67}$ Uppsala University, Box 516, SE-75120 Uppsala, Sweden

% ...
$^{c}$ Also at the Novosibirsk State University, Novosibirsk, 630090, Russia

% ...

}
 }

\abstract{Using a total of $5.25~{\rm fb}^{-1}$ of $e^+e^-$ collision data with center-of-mass energies from 4.236 to 4.600 GeV, we report the first observation of the process $\EE\to \eta\psip$ with a statistical significance of $5\sigma$.  The data sets were collected by the BESIII detector operating at the BEPCII storage ring.  We measure the yield of events integrated over center-of-mass energies and also present the energy dependence of the measured cross section.}


\begin{document} 

\maketitle
\flushbottom

\section{Introduction}
\label{sec:intro}

The recent observation of a number of unexpected vector charmoniumlike  states ($J^{PC}=1^{--}$) above open-charm threshold has stimulated theoretical and experimental studies of the
conventional and exotic states in this energy region~\cite{theory-Y-states-chenhuaxing-2016, theory-Y-states-Esposito-2017, theory-Y-states-Richard-2017, theory-Y-states-Ali-2017,  theory-Y-states-Stephen-2018,  theory-Y-states-guofenghun-2018, theory-Y-states-Brambilla-2020}.
These vector states, originally called
the $Y(4260)$~\cite{intro-BaBar-Y4260, intro-BaBar-Y4260-2012, intro-Belle-Y4260, intro-Belle-Y4260-2, intro-CLEO-Y4260-1},
the $Y(4360)$~\cite{intro-BaBar-Y4360, intro-BaBar-Y4360-Y4660-2014, intro-Belle-Y4360-Y4660},
and the $Y(4660)$~\cite{intro-BaBar-Y4360-Y4660-2014, intro-Belle-Y4360-Y4660},
first observed by the BaBar, Belle, and CLEO experiments, 
can be produced  via the initial state radiation (ISR) process,  and are often observed in final states with two pions and a state of charmonium, like the $\jpsi$ or $\psip$. 
They differ from the $\psi(3770)$, $\psi(4040)$, $\psi(4160)$, and $\psi(4415)$ states, which are well established experimentally in the $\EE$ inclusive hadronic cross section~\cite{pdg} and match potential model calculations of the charmonium spectrum~\cite{chao2}.
The $Y$ states have many theoretical interpretations, including compact tetraquarks, molecules,
hybrids, or hadrocharmonia~\cite{theory-Y-states-chenhuaxing-2016, theory-Y-states-Esposito-2017, theory-Y-states-Richard-2017, theory-Y-states-Ali-2017, theory-Y-states-Stephen-2018,  theory-Y-states-guofenghun-2018, theory-Y-states-Brambilla-2020}, and so on, but they still are mysterious.

In recent years two resonant structures around 4.22 and 4.32~GeV/$c^2$ were observed in a fit to the cross section of $\ee\to \ppjpsi$ measured by the BESIII
experiment~\cite{BESIII-pipijpsi-2017}. The lower mass structure, the $Y(4220)$, is the main component of the well-known $Y(4260)$ structure,
and the higher mass structure, the $Y(4320)$, could be the $Y(4360)$ resonance~\cite{babar-pipipsip-2014,intro-Belle-Y4360-2} observed in
the process $\ee\to \pppsip$. A series of cross section measurements of $\ee\to \pphc$~\cite{bes3-Y4230-pipihc}, $\ee\to \omega\chi_{c0}$~\cite{bes3-Y4230-omegachic0},
and $\ee\to \pip D^0D^{*-}+c.c.$~\cite{intro-Y4220-bes3-open-charm} has been reported by BESIII, and the parameters of the $Y(4220)$ resonance in these processes are consistent with those measured in the $\ee\to\ppjpsi$ process~\cite{BESIII-pipijpsi-2017}.

Searching for new decay modes of $Y$ states produced in $\EE$ annihilation and measuring the line shapes of the production cross sections will shed light on the nature of the $Y$ states. Besides the $\pi\pi$ hadronic transitions, other hadronic transitions (via $\eta$, $\etap$) of these $Y$ states to
lower mass charmonia such as the $\jpsi$ or $\psip$ also provide further insight into their internal structure.
The CLEO-c~\cite{intro-CLEO-etajpsi}, Belle~\cite{intro-Belle-etajpsi}, and BESIII~\cite{intro-Bes3-etajpsi, intro-Bes3-etajpsi-2, intro-Bes3-etajpsi-3} experiments measured the cross section of $\EE\to \eta\jpsi$, and BESIII observed the decays of the $Y(4220)$ and $Y(4390)$ into $\eta\jpsi$ final states. The authors of ref.~\cite{theory-etapsi-etapjpsi} reproduced the measured $\EE\to \eta\jpsi$ line shape and predicted the production cross section of the analogous process $\EE\to \etap\jpsi$ at ${\cal O}(\alpha_s^4)$ accuracy in the framework of nonrelativistic Quantum Chromodynamics (NRQCD). However, the measured cross sections of $\EE\to \etap\jpsi$~\cite{intro-bes3-etapjpsi, intro-bes3-etapjpsi2} by BESIII
are significantly smaller than the theoretical prediction~\cite{theory-etapsi-etapjpsi}.

To provide more information to study the vector charmonium(like) states, the cross section of  $\ee\to\eta\psip$ can also be compared with those of the processes $\EE\to \eta\jpsi$ and $\EE\to\etap\jpsi$. The CLEO-c experiment searched for the process $\EE\to \eta\psip$ with data at center-of-mass (c.m.) energy $\sqrt{s}=4.260$~GeV, and reported an upper limit on the Born cross section, $\sigma[\EE\to \eta\psip]<25$~pb, at a 90\% confidence level (C.L.)~\cite{intro-CLEO-etajpsi}. This is the only available experimental study of this process.

In this article, we present a study of $\EE\to \eta\psip$ at 14 c.m.\ energies from 4.236 to 4.600~GeV, using data collected with the BESIII detector~\cite{Ablikim-2009aa} operating at the BEPCII collider~\cite{Yu-IPAC2016-TUYA01}. The total integrated luminosity is 5.25~fb$^{-1}$. The c.m.\ energies were measured using $\EE\to \MM$ events with an uncertainty of 0.8~MeV~\cite{bes3-energy-measurement} and the integrated luminosities were measured using Bhabha scattering events to an uncertainty of 1.0\%~\cite{luminosity-measurement, luminosity-measurement-2}. The $\psip$ is reconstructed using the decay chain $\psip\to\ppjpsi$, $\jpsi \to \LL$ ($\ell = e$, $\mu$), and the $\eta$ using $\eta\to\gamma\gamma$.

\section{BESIII detector and Monte Carlo simulation}
\label{sec:detector}

The cylindrical core of the BESIII detector consists of a helium-based multilayer drift chamber (MDC), a plastic scintillator time-of-flight system (TOF), and a CsI(Tl) electromagnetic calorimeter (EMC), which are all enclosed in a superconducting solenoidal magnet providing a 1.0~T magnetic field. The solenoid is supported by an octagonal flux-return yoke with resistive plate chamber muon identifier modules interleaved with steel. The acceptance of charged particles and photons is 93\% over $4\pi$ solid angle. The charged-particle momentum resolution at $1~{\rm GeV}/c$ is $0.5\%$, and the $dE/dx$ resolution is $6\%$ for the electrons from Bhabha scattering events. The EMC measures photon energies with a resolution of $2.5\%$ ($5\%$) at $1$~GeV in the barrel (end cap) region. The time resolution of the TOF barrel part is 68~ps, while that of the end cap part is 110~ps. The end cap TOF system was upgraded in 2015 with multi-gap resistive plate chamber technology, providing a time resolution of 60~ps~\cite{etof, etof-2}.

To optimize the signal event selection criteria, estimate the background contributions and determine the detection efficiency, simulated samples are produced with the {\sc geant4}-based~\cite{geant4} Monte Carlo (MC) package which includes the geometric description of the BESIII detector and the detector response. The signal MC events of $\EE\to \eta\psip$ with the corresponding $\eta$ and $\psip$ decay modes are generated using {\sc HELAMP}  and {\sc evtgen}~\cite{ref-evtgen, ref-evtgen-2} at each c.m.\ energy. The beam energy spread and ISR in the $\EE$ annihilations are modelled with the generator {\sc kkmc}~\cite{ref-kkmc, ref-kkmc-2}  and the final state radiations (FSR) from charged final-state particles are incorporated with the {\sc photos} package~\cite{photos}. The possible background contributions are also studied with  {\sc kkmc}~\cite{ref-kkmc, ref-kkmc-2} at each c.m.\ energy. The decay modes are modelled with {\sc evtgen}  using branching fractions taken from the PDG~\cite{pdg}.

\section{Event selection}
\label{sec:eventselection}

Candidate events with four charged tracks with zero net charge and at least two photons are selected. The charged tracks are required to be well reconstructed in the MDC with a polar angle $\theta$ satisfying $|\cos\theta|<0.93$, and the distances of the closest approach to the interaction point in $x-y$ plane and $z$ direction have to be  less than 1~cm and 10~cm, respectively. Since the $\pi^{\pm}$ and $\ell^{\pm}$ are kinematically well separated, charged particles with momenta less than $0.8~\gev/c$ in the laboratory frame are assumed to be $\pi^{\pm}$, whereas the ones with momenta larger than $1.0~\gev/c$ are assumed to be $\ell^{\pm}$. To separate electron from muon candidates, the EMC deposited energy is used. The energy deposits of electron candidates and muon candidates are required to be larger than $1.0~\gev$ and less than $0.4~\gev$, respectively. Photon candidates are reconstructed from showers in EMC crystals. The reconstructed energies for the clusters in the barrel ($|\cos{\rm \theta|}<0.80$) and the end caps ($0.86<|\cos{\rm \theta}|<0.92$) of the EMC are required to be higher than 25 and 50~MeV, respectively. To eliminate showers associated with charged particles, the angle between the photon and any charged track in the EMC must be at least 10 degrees. To suppress the electronic noise and energy deposits unrelated to the event, the time of the EMC shower is required to be $0\leqslant t \leqslant 700$~ns with respect to the start of the event. To improve the mass resolution and suppress background contributions, a four-constraint~(4C) kinematic fit is performed under the hypothesis of $\ee\to\gamma\gamma\pip\pim \ell^{+}\ell^{-}$  to constrain the sum of four momenta of the final state particles to the initial colliding beams. The chi-square of the kinematic fit, $\chi^{2}_{\rm 4C}$, is required to be less than 40. If there are more than two photons in an event, the combination of $\gamma\gamma\pip\pim \ell^{+}\ell^{-}$ with the least $\chi^{2}_{\rm 4C}$  is retained for further study.

\begin{figure*}[tbp]
\centering
\subfigure{
\label{fig:2d-mass-eta-pipijpsi-a}
\includegraphics[width=0.22\paperwidth]
 {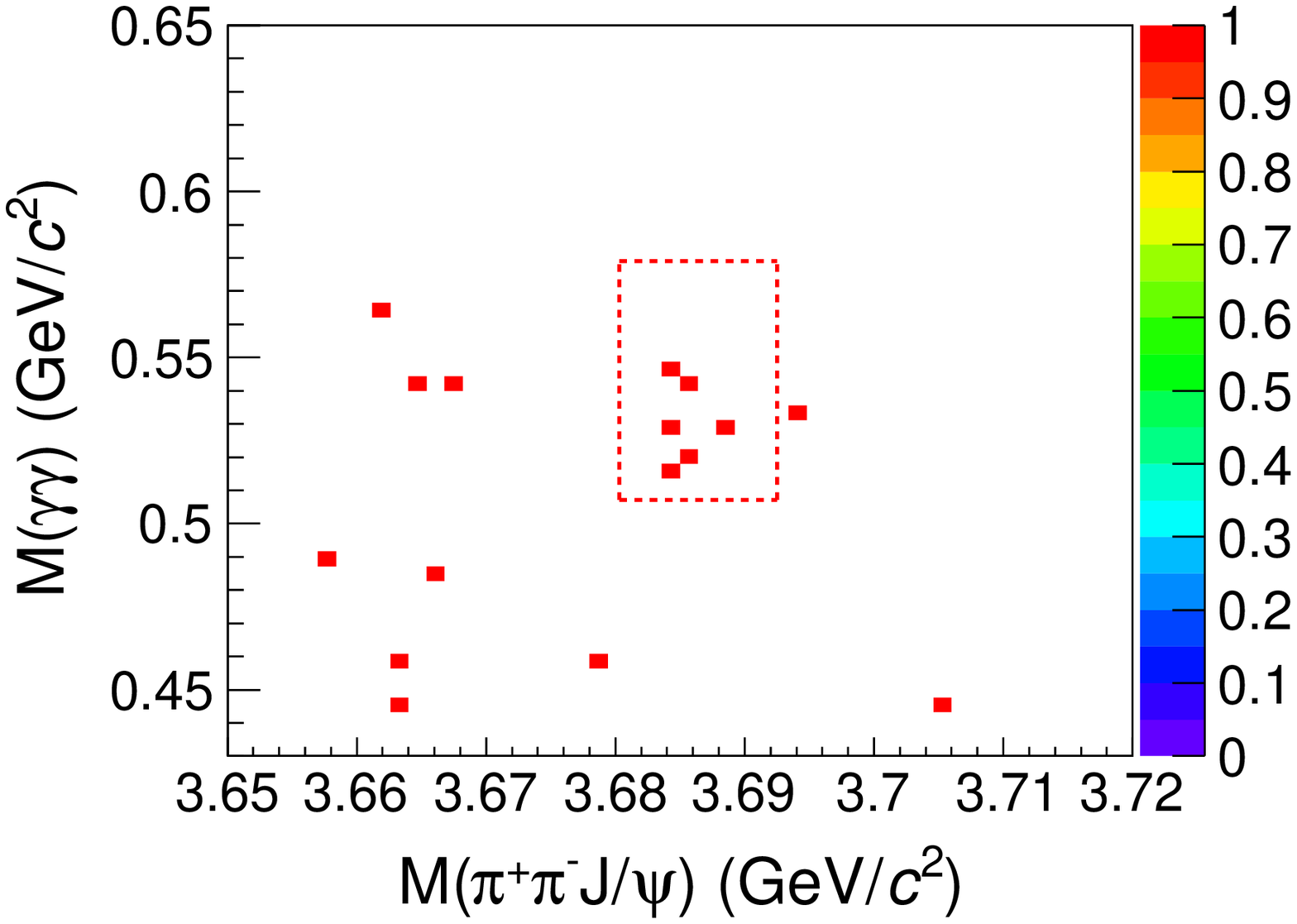}
\put(-100,75){\textbf{(a)}}
 }
\subfigure{
 \label{fig:2d-mass-eta-pipijpsi-e}
\includegraphics[width=0.22\paperwidth]
 {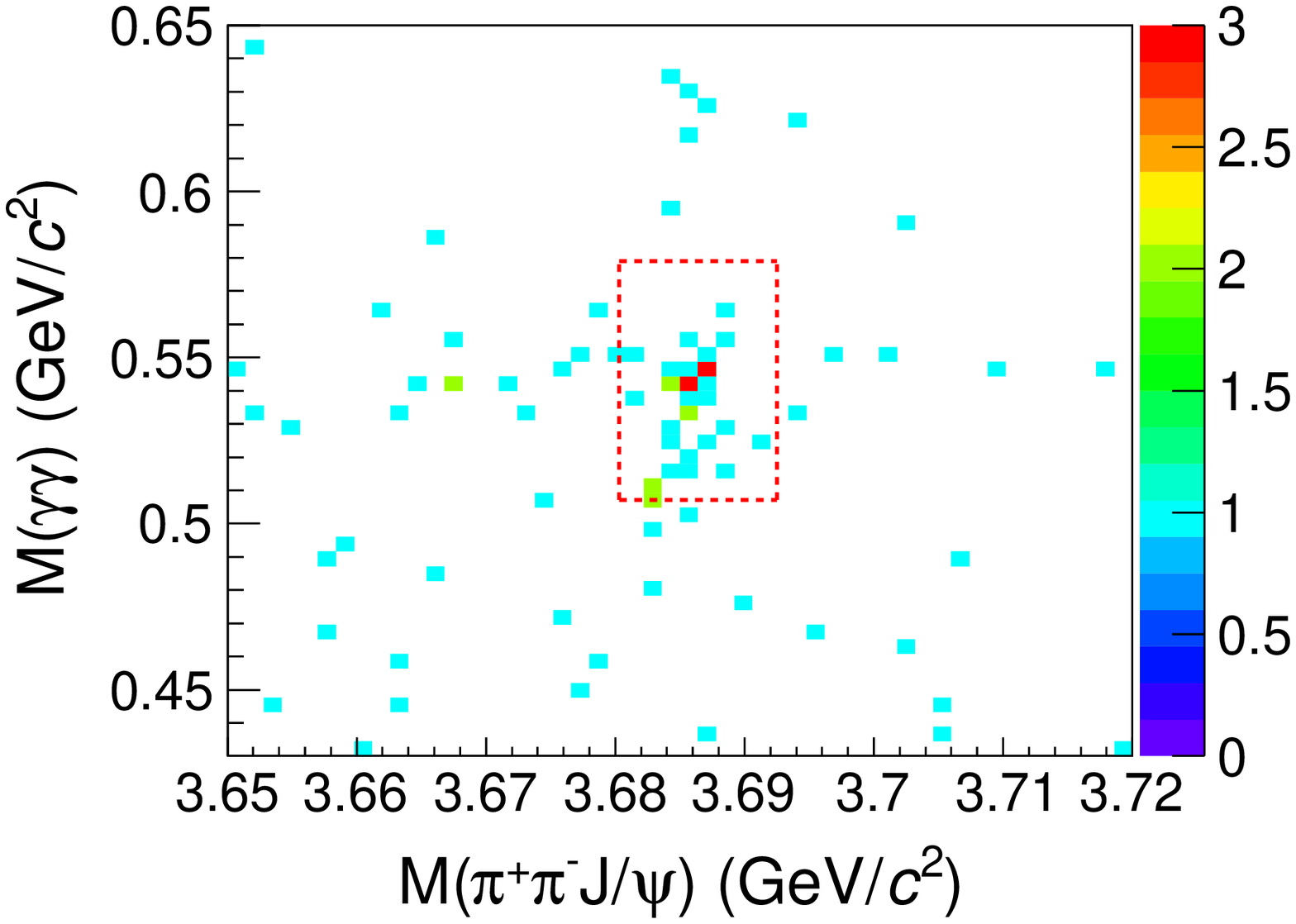}
\put(-100,75){\textbf{(f)}}
 }\\
 \subfigure{
\label{fig:2d-mass-eta-pipijpsi-b}
\includegraphics[width=0.22\paperwidth]
 {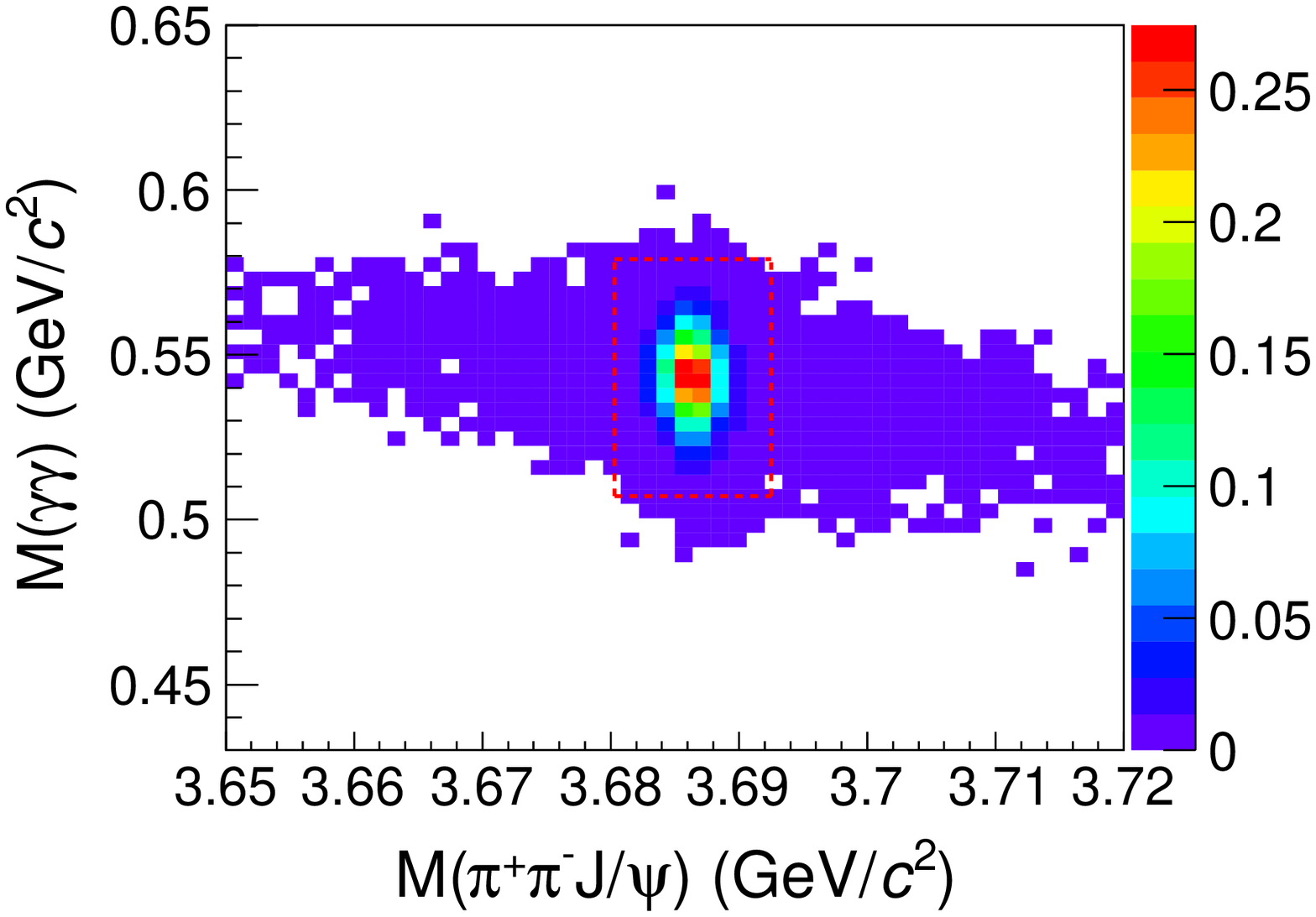}
\put(-100,75){\textbf{(b)}}
 }
 \subfigure{
\label{fig:2d-mass-eta-pipijpsi-f}
\includegraphics[width=0.22\paperwidth]
 {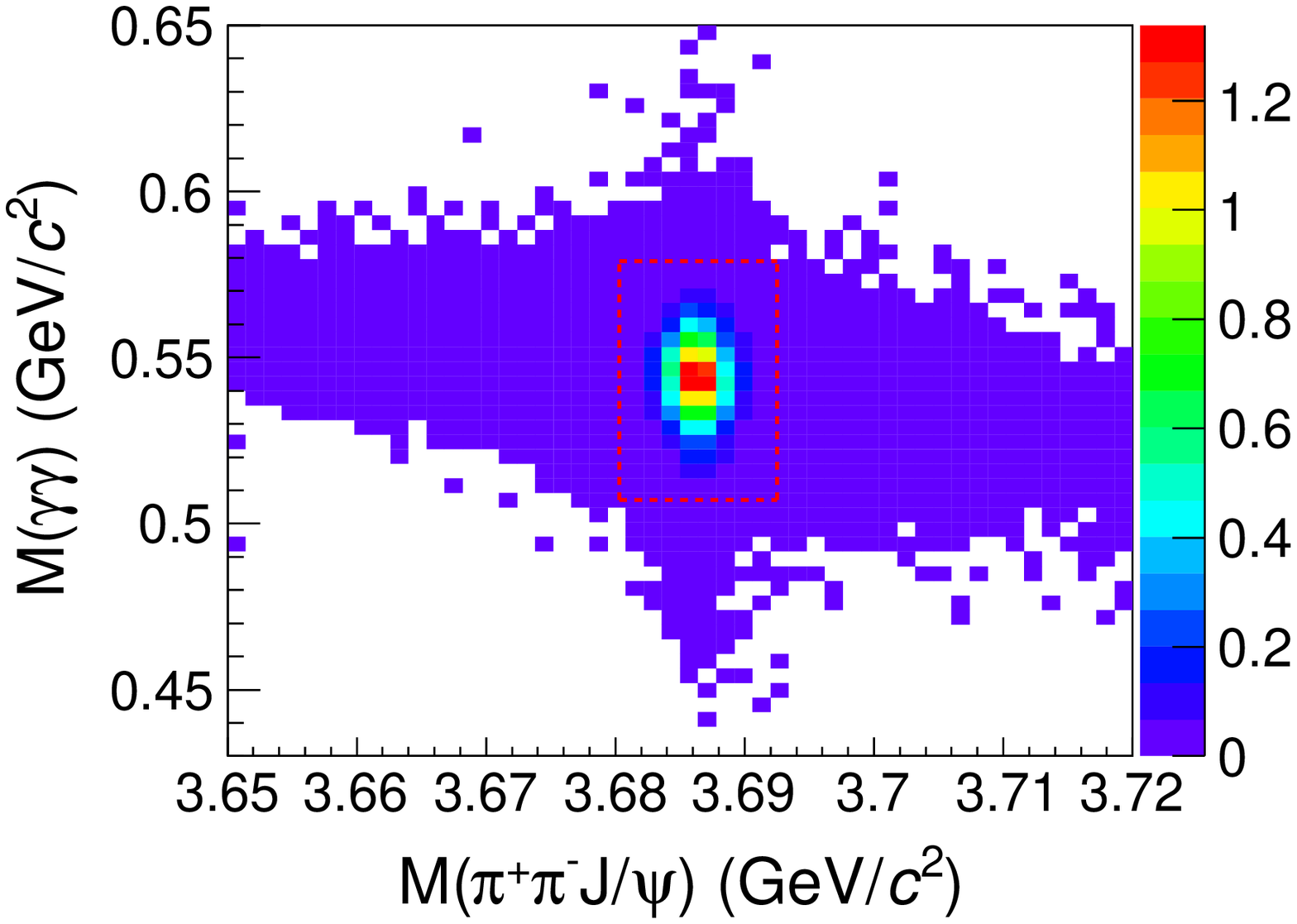}
\put(-100,75){\textbf{(g)}}
 }\\
  \subfigure{
\label{fig:2d-mass-eta-pipijpsi-b}
\includegraphics[width=0.22\paperwidth]
 {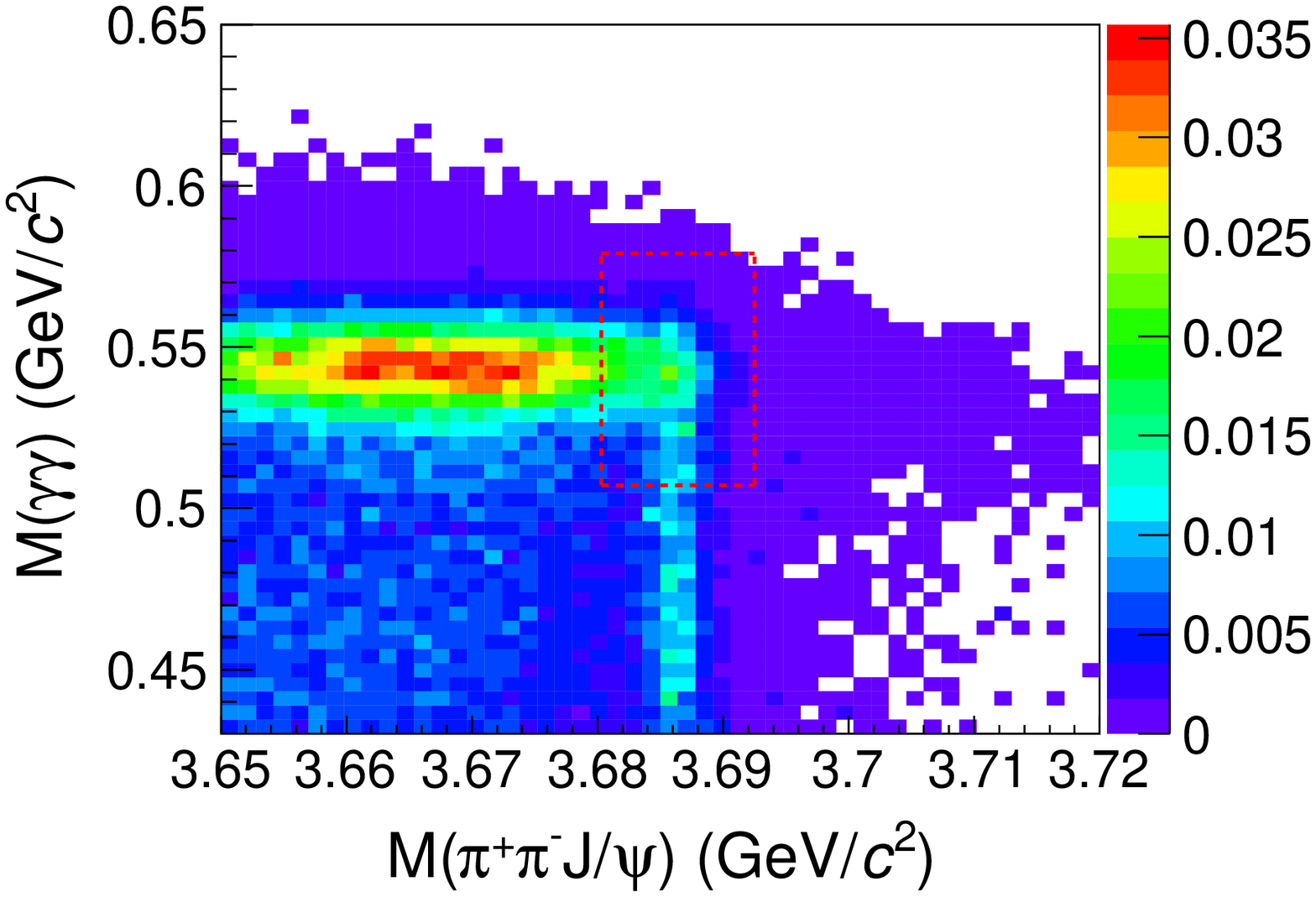}
\put(-100,75){\textbf{(c)}}
 }
    \subfigure{
\label{fig:2d-mass-eta-pipijpsi-f}
\includegraphics[width=0.22\paperwidth]
 {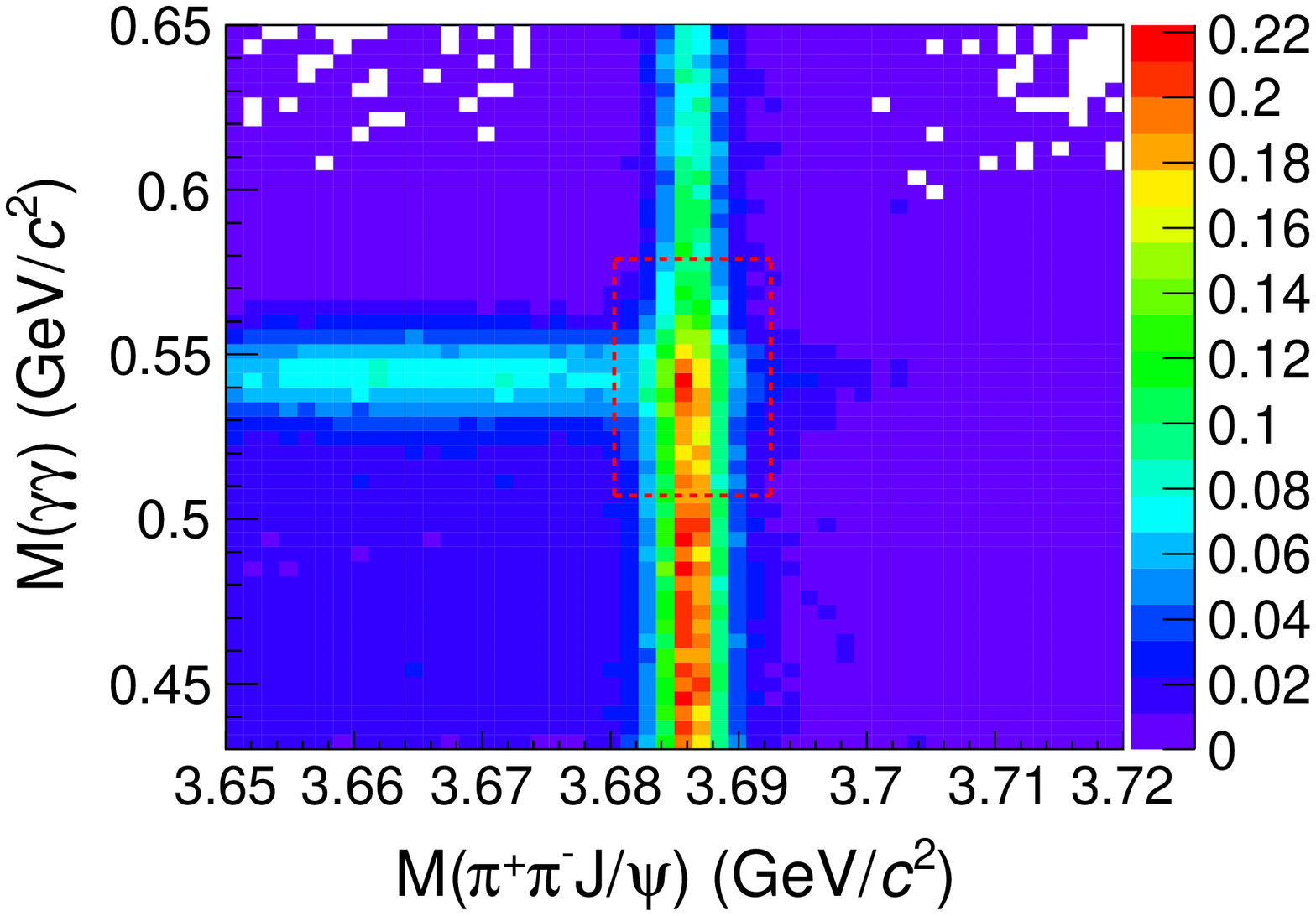}
\put(-100,75){\textbf{(h)}}
 }\\
 \subfigure{
 \label{fig:2d-mass-eta-pipijpsi-c}
\includegraphics[width=0.22\paperwidth]
 {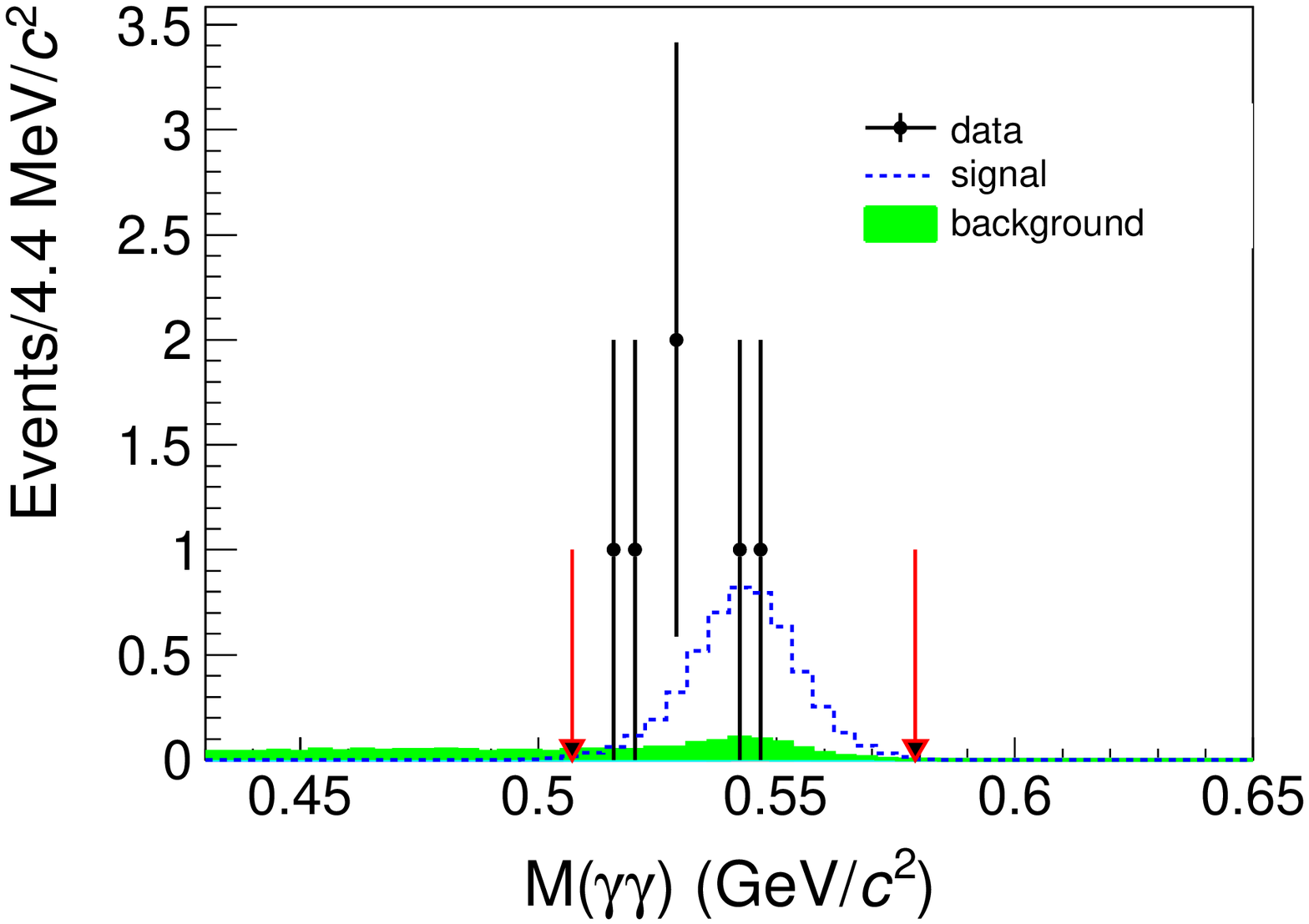}
\put(-100,75){\textbf{(d)}}
 }
  \subfigure{
 \label{fig:2d-mass-eta-pipijpsi-g}
\includegraphics[width=0.22\paperwidth]
 {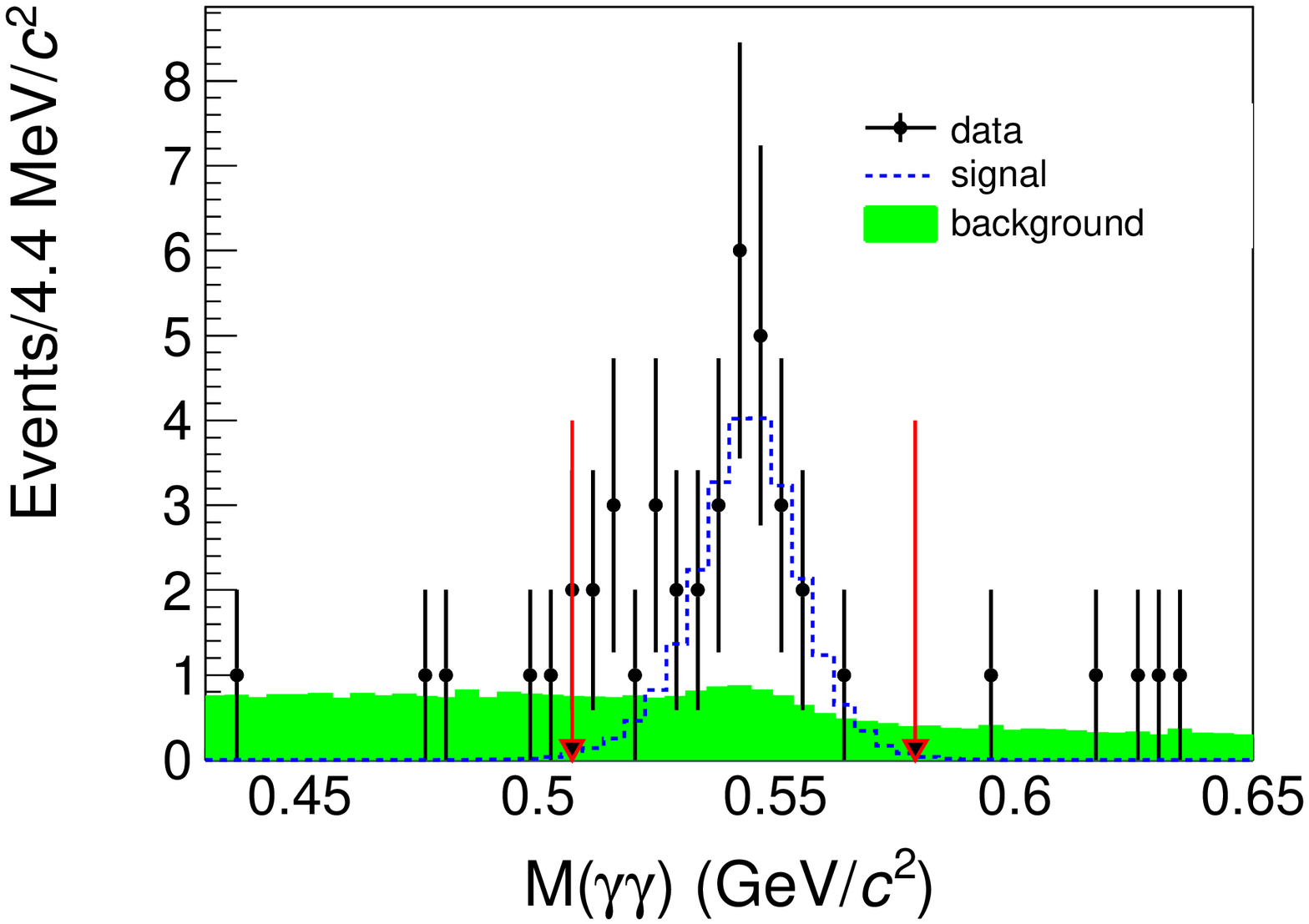}
\put(-100,75){\textbf{(i)}}
 }\\
 \subfigure{
 \label{fig:2d-mass-eta-pipijpsi-d}
\includegraphics[width=0.22\paperwidth]
 {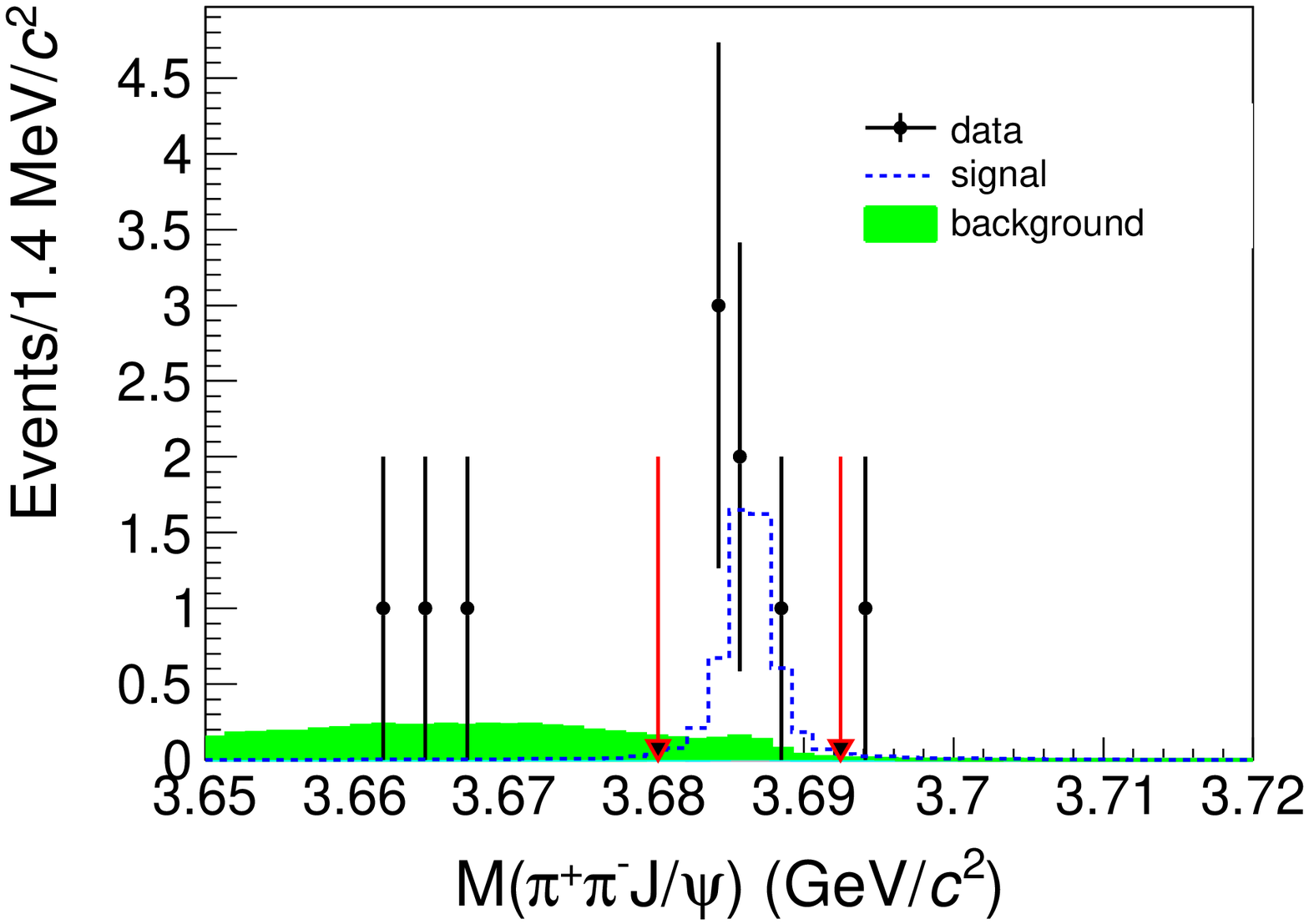}
\put(-100,75){\textbf{(e)}}
 }
 \subfigure{
 \label{fig:2d-mass-eta-pipijpsi-h}
\includegraphics[width=0.22\paperwidth]
 {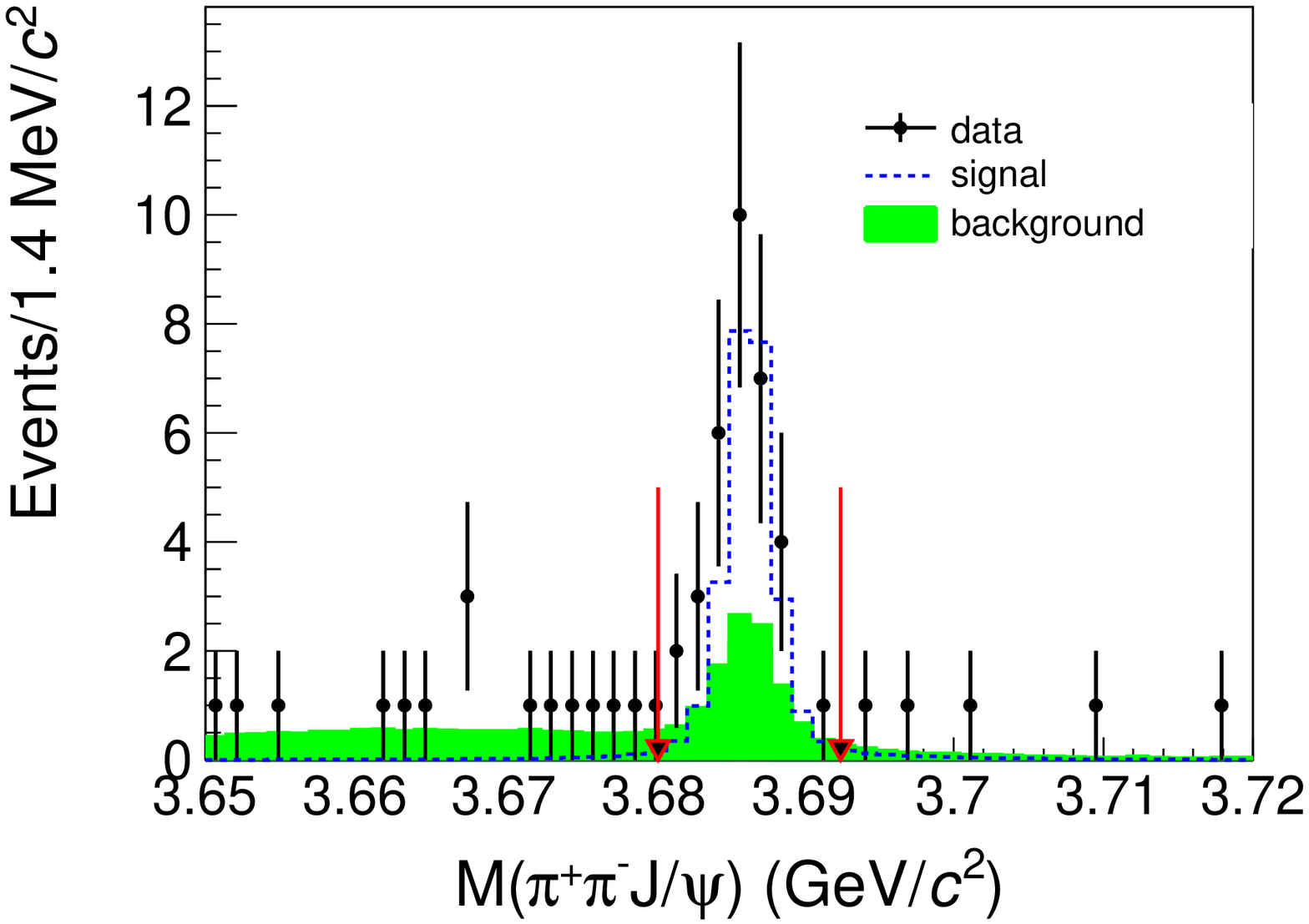}
\put(-100,75){\textbf{(j)}}
 }
\caption{\label{fig:2d-mass-eta-pipijpsi}
Two-dimensional distributions of $M(\gamma\gamma)$ versus M($\pip\pim \jpsi$) for (a) data,  (b) signal MC simulation,  and (c) background MC contributions with the red dashed boxes for the defined $\eta$ and $\psip$ signal regions, and the corresponding projections  of (d) $M(\gamma\gamma)$ distribution in the $\psip$ mass window and (e) $M(\pip\pim \jpsi)$ distribution in the $\eta$ mass window with red arrows for the defined signal regions at $\sqrt{s}=4.258$ GeV, where the dots with error bars, the dashed blue lines, and the green histograms represent  data, signal MC, and background MC simulations, respectvely. The same distributions for the sum of the 14 data samples and MC samples are shown in (f), (g), (h),  (i), and (j) correspondingly.}
\end{figure*}

To identify signal candidates that involve the $\jpsi$ resonance, we select events with a $\LL$ invariant mass within a window of  $\pm3\sigma$ around the $\jpsi$ nominal mass,  $3064.6 < M(\LL)< 3140.8$~MeV/$c^2$, referred to as the $\jpsi$ mass window.  To remove the background from process $\EE \to \etap\jpsi$ with $\etap\to \pip\pim\eta$, the invariant mass of $\pip \pim \gamma\gamma$ is required to be larger than $1~\gevcc$. Two-dimensional~(2D) distributions for $\gamma\gamma$ and $\ppjpsi$ invariant masses, $M(\gamma\gamma)$ versus $M(\ppjpsi)$, and the corresponding one-dimensional~(1D) projections for data, signal MC samples, background contributions at $\sqrt{s}=4.258$~GeV are presented in figures~\ref{fig:2d-mass-eta-pipijpsi}(a-e). The distributions for the sum of 14 energy points are shown in figures~\ref{fig:2d-mass-eta-pipijpsi}(f-j). Signal candidates are required to be within the  $\eta$ mass region [$\pm3\sigma$ around the $\eta$ nominal mass], defined as $507.1<M(\gamma\gamma)<579.1$~MeV/$c^2$, and  $\psip$ mass region  [$\pm3\sigma$ around the $\psip$ nominal mass], defined as   $3680.3<M(\ppjpsi)<3692.5$~MeV/$c^2$  (as indicated by red dashed boxes or the ranges between two arrows in figure~\ref{fig:2d-mass-eta-pipijpsi}). Significant clusters can be seen in the mass windows of the $\eta$ and $\psip$.

\section{Background analysis}
\label{sec:background}
To study background processes, we generated a series of MC samples for final states that include a $\pip\pim$ pair, two leptons with high momenta, and at least two photons in the final state using the {\sc kkmc} generator at each energy point. These background processes are listed in table~\ref{table:The possible backgrounds}. The dominant background contribution is $\EE \to \gamma\gamma\psip$, and it is measured directly in this analysis.  The yields for each of the other background processes in the 2D signal region ($N_{{\rm bkg},i}$) are calculated using external input by:
\begin{equation}\label{equ:bk number}
\begin{split}
\begin{aligned}
 N_{{\rm bkg},i}=\mathcal{L}_{\rm int}(1+\delta)_i |1-\Pi|^{-2}
\epsilon_i\mathcal{B}_i\sigma^{\rm B}_{{\rm bkg},i},
  \end{aligned}
\end{split}
\end{equation}
where $i$ represents each background channel, $\mathcal{L}_{\rm int}$ is the integrated luminosity, $|1-\Pi|^{-2}$ is the vacuum polarization factor~\cite{ vacuum-polarization-factor}, $\epsilon_i$ and $\mathcal{B}_i$ are the selection efficiency and the product branching fraction of the intermediate states taken from
the PDG~\cite{pdg} for the $i$th background mode, respectively, and $\sigma^{\rm B}_{{\rm bkg},i}$ is the measured Born cross section of the $i$th background mode.
The production cross sections for these background processes are  taken from refs.~\cite{cross-section-pppsip,cross-section-p0p0psip,cross-section-omegachic012, cross-section-omegachic012-2, bes3-Y4230-omegachic0,cross-section-gammax3872,cross-section-phichic12}.  
$(1+\delta)_i$ is the ISR correction factor obtained from a quantum electrodynamics calculation~\cite{ref-kkmc, ref-kkmc-2, isr-calculate2} using the {\sc kkmc} generator, assuming an input lineshape from refs.~\cite{cross-section-pppsip,cross-section-p0p0psip,cross-section-omegachic012, cross-section-omegachic012-2, bes3-Y4230-omegachic0,cross-section-gammax3872,cross-section-phichic12}.

\begin{table*}[tbp]
\centering
\begin{tabular}{|lll|}
\hline 
\multicolumn{3}{|c|}{Decay mode}\\\hline
$\EE \to \pip\pim\psip$,       &$\psip\to \jpsi \eta$,             & $\eta \to \gamma \gamma$ \\ 
$\EE \to \pip\pim\psip$,       &$\psip\to \gamma \chi_{cJ}(J=0,1,2)$ ,   & $\chi_{cJ} \to \gamma \jpsi$ \\
$\EE \to \pi^{0}\pi^{0}\psip$, &$\psip  \to \ppjpsi$               &  \\ 
$\EE \to \omega  \chi_{cJ}(J=0,1,2)$, &$\omega \to\pip\pim\pi^{0}$,     & $\chi_{cJ} \to \gamma \jpsi$ \\  
$\EE \to \gamma X(3872)$,        &$X(3872) \to \omega \jpsi$,      & $\omega \to \pip\pim\pi^{0}$ \\ 
$\EE \to \phi \chi_{cJ}(J=1,2)$,      &$\phi \to \pip \pim \pi^{0}$,    & $\chi_{cJ} \to \gamma \jpsi$   \\
$\EE \to \gamma\gamma\psip$, &$\psip  \to \ppjpsi$                 &  \\ \hline
\end{tabular}
\caption{The background processes~(all $\jpsi$ mesons decay into $\LL$).}\label{table:The possible backgrounds}
\end{table*}

The irreducible background process $\EE \to \gamma \gamma \psip$, $\psip\to\ppjpsi$, $\jpsi \to \LL$ with the two photons not from resonance decay has the same final
state particles as the signal channel, thus we measure its yield with the data directly. After applying all the selection criteria as for signal but the $\eta$ mass window,
we veto  processes from $\ee\to \gamma_{\rm ISR}\psip$, $\ee\to \pi^{0}\psip$, and $\ee\to \eta\psip$ by requiring the mass range of $\gamma\gamma$ larger than 300~MeV/$c^2$
and not in $[507.1, 579.1]$~MeV/$c^{2}$. We fit $M(\pip\pim \jpsi)$ distribution of data using the line shape of MC simulated $\EE\to \gamma \gamma \psip$ events to obtain the number of $\ee\to \gamma\gamma\psip$ events [$N_{\gamma\gamma\psip}^0$] at each c.m.\ energy. The number of $\ee\to \gamma\gamma\psip$ events in the $\eta$ and $\ppjpsi$ signal regions [$N_{\gamma\gamma\psip}^1$] is  obtained from the $N_{\gamma \gamma \psi(2S)}^{0}$ as follows:
\begin{equation}
N_{\gamma\gamma\psip}^1 = F\cdot N_{\gamma \gamma \psi(2S)}^{0},
\end{equation}
\begin{equation}
F = \frac{\epsilon^{1}_{e}\mathcal{B}_{e}+\epsilon_{\mu}^{1}\mathcal{B}_\mu}{\epsilon^{0}_{e}\mathcal{B}_{e}+\epsilon_{\mu}^{0}\mathcal{B}_\mu},
\end{equation}
where $F$ is a factor constructed from branching fractions and selection efficiencies,   $\epsilon^{1}_{e}$ and $\epsilon_{\mu}^{1}$ are the detection efficiencies for
$\jpsi\to\EE$ and $\jpsi\to\MM$ decay channels in the $\eta$ and $\ppjpsi$ signal regions, respectively; $\epsilon^{0}_{e}$ and $\epsilon_{\mu}^{0}$ are those of efficiencies outside the $\eta$  signal region;  $\mathcal{B}_e$ and $\mathcal{B}_\mu$ are the branching fractions of decays  $\jpsi\to\EE$, and $\jpsi\to\MM$, respectively~\cite{pdg}.

The number of $\ee\to \gamma\gamma\psip$ events outside the $\eta$  signal region [$N_{\gamma \gamma \psi(2S)}^{0}$] and the $F$ factor at each c.m.\ energy are listed in table~\ref{table:results for gamma gamma psip}. 
\begin{table}[tbp]
\centering
\begin{tabular}{|c|c|c|}
\hline
$\sqrt s$~(GeV) & $N_{\rm \gamma\gamma\psip}^0$
& $F$  \\\hline
   4.236 &  $ 5.94^{+3.12}_{-2.40}$ &   0.106 \\
   4.242 &  $ 1.79^{+1.80}_{-1.16}$ &   0.125 \\
   4.244 &  $ 5.99^{+2.80}_{-2.12}$ &   0.130 \\
   4.258 &  $ 1.35^{+2.49}_{-1.35}$ &   0.184 \\
   4.267 &  $ 1.81^{+2.13}_{-1.40}$ &   0.210 \\
   4.278 &  $ 2.48^{+2.53}_{-1.81}$ &   0.239 \\
   4.308 &  $ 0.00^{+1.29}_{-0.00}$  &   0.261 \\
   4.358 &  $ 5.04^{+3.12}_{-2.44}$ &   0.255 \\
   4.387 &  $ 0.00^{+1.29}_{-0.00}$  &   0.239 \\
   4.416 &  $11.28^{+4.49}_{-3.79}$ &   0.228 \\
   4.467 &  $ 2.25^{+2.12}_{-1.34}$ &   0.197 \\
   4.527 &  $ 0.00^{+1.29}_{-0.00}$  &   0.176 \\
   4.575 &  $ 1.00^{+1.36}_{-0.70}$ &   0.159 \\
   4.600 &  $ 4.02^{+2.97}_{-2.34}$ &   0.147 \\
  \hline
\end{tabular}
\caption{The number of $\ee\to \gamma\gamma\psip$ events outside the $\eta$  signal region [$N_{\gamma \gamma \psi(2S)}^{0}$] and the $F$ factor at each c.m.\ energy.}\label{table:results for gamma gamma psip}
\end{table}
The total number of background events ($n^{\rm b}$)  in the 2D signal region is obtained with
\begin{equation}\label{equ:bk number2}
\begin{split}
\begin{aligned}
 n^{\rm b}&=\sum_i N_{{\rm bkg},i}+N_{{\gamma\gamma\psip}}^1  .\\
   \end{aligned}
\end{split}
\end{equation}
Finally, the total number of background events in the signal region at different energy points, together with the number of background events from different
final states are listed in table~\ref{table:The expected events from different background MC samples}.

\begin{sidewaystable*}[tbp]
\centering
\begin{tabular}{|cccccccc|}
\hline
$\sqrt s$ (GeV)                        &4.236           &4.242              &4.244           &4.258           &4.267           &4.278            &4.308                                     \\\hline
$\pip\pim\psip,\psip\to\jpsi\eta$      &0.00           &0.00$\pm$0.00    &0.01           &0.49$\pm$0.05 &0.87           &0.39            &0.03$\pm$0.01                           \\
$\pip\pim\psip,\psip\to\gamma\chi_{c0}$&0.00           &0.00$\pm$0.00    &0.00           &0.01$\pm$0.00 &0.02           &0.01            &0.00                   \\
$\pip\pim\psip,\psip\to\gamma\chi_{c1}$&0.00           &0.00$\pm$0.00    &0.00           &0.08$\pm$0.01 &0.13           &0.05            &0.01$\pm$0.00         \\
$\pip\pim\psip,\psip\to\gamma\chi_{c2}$&0.00           &0.00$\pm$0.00    &0.00           &0.00$\pm$0.00 &0.00           &0.00            &0.00$\pm$0.00         \\
$\pi^{0}\pi^{0}\psip$                  &0.00           &0.00$\pm$0.00    &0.00$\pm$0.02 &0.04$\pm$0.01 &0.02$\pm$0.00 &0.01$\pm$0.00  &0.00$\pm$0.01         \\
$\omega \chi_{c0}$                     &0.00$\pm$0.00 &0.00$\pm$0.00    &0.00$\pm$0.00 &0.00$\pm$0.00 &0.00$\pm$0.00 &0.00$\pm$0.00  &0.00$\pm$0.00         \\
$\omega \chi_{c1}$                     &---             &---                &---             &---             &---             &---              &0.00$\pm$0.01         \\
$\omega \chi_{c2}$                     &---             &---                &---             &---             &---             &---              &---                              \\
$\gamma X(3872)$                       &---             &---                &---             &---             &---             &---              &---                              \\
$\phi \chi_{c1}$                       &---             &---                &---             &---             &---             &---              &---                              \\
$\phi \chi_{c2}$                       &---             &---                &---             &---             &---             &---              &---                              \\
$\gamma\gamma\psip$                    &$0.63_{-0.25}^{+0.33}$ &$0.22_{-0.15}^{+0.23}$&$0.78_{-0.27}^{+0.36}$&$0.25_{-0.25}^{+0.46}$&$0.38_{-0.29}^{+0.45}$&$0.59_{-0.43}^{+0.60}$&$0.00_{-0.00}^{+0.34}$  \\\hline
$n^{\rm b}$      &0.63$\pm$0.33           &0.22$\pm$0.23              &0.79$\pm$0.36           &0.88$\pm$0.46           &1.41$\pm$0.45           &1.06$\pm$0.60            &0.04$\pm$0.34                   \\\hline\hline
$\sqrt s$ (GeV)                        &4.358           &4.387                   &4.416              &4.467                    &4.527                      &4.575              &4.600             \\\hline
    $\pip\pim\psip,\psip\to\jpsi\eta$      &0.00$\pm$0.00 &0.00$\pm$0.00         &0.00$\pm$0.00    &---                  &---                        &---                                 &---               \\
$\pip\pim\psip,\psip\to\gamma\chi_{c0}$&0.00$\pm$0.00 &0.00$\pm$0.00         &0.00$\pm$0.00    &---                      &---                        &---                             &---               \\
$\pip\pim\psip,\psip\to\gamma\chi_{c1}$&0.00$\pm$0.00 &0.00$\pm$0.00         &0.00$\pm$0.00    &---                      &---                        &---                             &---               \\
$\pip\pim\psip,\psip\to\gamma\chi_{c2}$&0.00$\pm$0.00 &0.00$\pm$0.00         &0.00$\pm$0.00    &---                      &---                        &---                             &---               \\
$\pi^{0}\pi^{0}\psip$                  &0.20$\pm$0.03 &0.00$\pm$0.03         &0.29$\pm$0.04    &0.00$\pm$0.02          &0.00$\pm$0.02            &0.00$\pm$0.01    &0.02$\pm$0.01   \\
$\omega \chi_{c0}$                     &0.00$\pm$0.00 &0.00$\pm$0.00         &0.00$\pm$0.00    &0.00$\pm$0.00          &0.00$\pm$0.00            &0.00$\pm$0.00    &0.00$\pm$0.00   \\
$\omega \chi_{c1}$                     &0.00$\pm$0.01 &0.00$\pm$0.01         &0.00$\pm$0.04    &0.00$\pm$0.01          &0.00$\pm$0.00            &---                             &---               \\
$\omega \chi_{c2}$                     &0.00$\pm$0.05 &0.00$\pm$0.02         &0.16$\pm$0.03    &0.00$\pm$0.02          &0.00$\pm$0.00            &0.00$\pm$0.00    &0.00$\pm$0.01   \\
$\gamma X(3872)$                       &0.00$\pm$0.00 &---                     &0.00$\pm$0.00    &---                      &---                        &---                &0.00$\pm$0.00   \\
$\phi \chi_{c1}$                       &---             &---                     &---                &---                      &---                        &---                &0.00$\pm$0.00   \\
$\phi \chi_{c2}$                       &---             &---                     &---                &---                      &---                        &---                &0.00$\pm$0.00   \\
$\gamma\gamma\psip$                    &$1.28_{-0.62}^{+0.79}$&$0.00_{-0.00}^{+0.31}$&$2.57_{-0.86}^{+1.02}$&$0.44_{-0.26}^{+0.42}$&$0.00_{-0.00}^{+0.23}$&$0.16_{-0.11}^{+0.22}$&$0.59_{-0.34}^{+0.44}$\\\hline
$n^{\rm b}$      &1.49$\pm$0.80           &0.00$\pm$0.31                   &3.03$\pm$1.02              &0.44$\pm$0.42                    &0.00$\pm$0.23                      &0.16$\pm$0.22              &0.62$\pm$0.44           \\\hline
\end{tabular}
\caption{The total number of background events  in the signal region ($n^{\rm b}$) at different energy points, together with the number of background events from different final states. Ellipses mean that the results are not applicable, and the numbers which are less than 0.005 are indicated with 0.00.}\label{table:The expected events from different background MC samples}
\end{sidewaystable*}

\section{Cross section measurement}
\label{sec:cross}
It is assumed that the number of observed events ($n^{\rm obs}$) in the signal region follows a Poisson distribution, with the numbers of expected background
($n^{\rm  b}$) and signal ($\mu$) events, respectively,
\begin{equation}\label{equ:Poisson}
  P(n^{\rm obs};\mu, n^{\rm b})=(\mu+ n^{\rm b})^{n^{\rm obs}}e^{-(\mu+ n^{\rm b})}/n^{\rm obs}!.
\end{equation}
There are some energy points where the number of observed events is zero, but the number of background events is non-zero, such as $\sqrt{s}=4.244$~GeV.
Using the same method as in ref.~\cite{feldman-cousins-method}, the value of $\mu$ with the maximum $P(n^{\rm obs};\mu, n^{\rm b})$ is taken as the non-negative number of signal events ($n^{\rm sig}$). Thus, $n^{\rm sig}$ = max(0, $n^{\rm obs}- n^{\rm b}$) is the best estimation of the number of signal events in the physically-allowed region.

The statistical uncertainty of the number of signal events at a 68.27\% C.L.\ is estimated with the Feldman-Cousins~(FC) method~\cite{feldman-cousins-method}. Since no significant $\eta\psip$ signal events are observed at some energy pints, the upper limits at a 90\% C.L.\ for the number of signal events are obtained with the Poissonian limit estimator (POLE) computer program~\cite{pole-method}.

\begin{figure}[tbp]
\centering
\includegraphics[width=0.4\paperwidth] {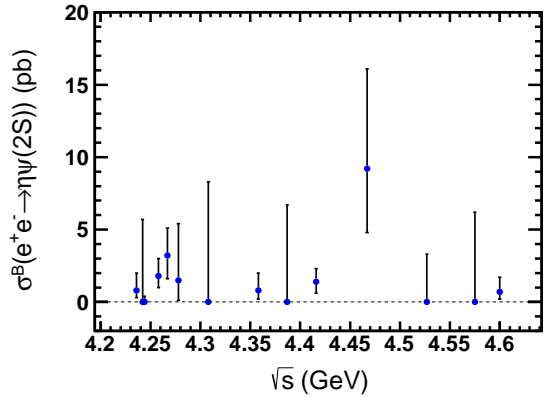}
\caption{\label{fig:cross section}The measured Born cross section as a function of the collision energy. The uncertainties are statistical only.}
\end{figure}

The Born cross section of $\EE\to\eta\psip$ is calculated with
\begin{equation}
\begin{split}
\begin{aligned}
 \sigma^{\rm B}=\frac{n^{\rm sig}}
 {\mathcal{L_{\rm int}}(1+\delta)|1-\Pi|^{-2}
   \mathcal{B}_1
   \mathcal{B}_2
  (\epsilon_e\mathcal{B}_e+\epsilon_\mu\mathcal{B}_\mu)
   },
  \end{aligned}
   \end{split}
\end{equation}
where $\mathcal{B}_1$ and $\mathcal{B}_2$ are the branching fractions of $\psip\to\ppjpsi$ and $\eta\to\gamma\gamma$~\cite{pdg}, respectively; $(1+\delta)$ is the radiative correction factor obtained from  the quantum electrodynamics calculation~\cite{ref-kkmc, ref-kkmc-2, isr-calculate2} using the {\sc kkmc} generator, assuming an input lineshape of the $Y(4260)$~\cite{pdg} cross section. The Born cross sections (and upper limits at the 90\% C.L.), and the numbers used in the calculation are listed in table~\ref{table:Preliminary results}.
\begin{sidewaystable*}[htbp]
\centering
\begin{tabular}{|ccccccccccc|}
\hline
$\sqrt s$~(GeV) &$ \mathcal{L}_{\rm int}$~(pb$^{-1}$)&$n^{\rm obs}$&$n^{\rm b}$ &$n^{\rm sig}$&$n^{\rm sig}_{\rm POLE}$
& $\Sigma(10^{-2})$ &$(1+\delta)$ &$|1-\Pi|^{-2}$ &$\sigma^{B}$~(pb)  & $\sigma^{\rm B}_{\rm POLE}$~(pb)\\\hline
        4.236 & 530.3 & 2 &$0.63\pm0.33$  & $1.4^{+2.2}_{-1.0}$&$ (0.0    ,   5.9 )$ &0.430  &0.76 &1.056 &$0.8^{+1.2}_{-0.5} $ &(0.0,~3.2) \\
        4.242 & 55.9  & 0 &$0.22\pm0.23$  & $0.0^{+1.1}_{-0.0}$&$ ( 0.0   ,    2.5)$ &0.430  &0.76 &1.055 &$0.0^{+5.7}_{-0.0} $ &(0.0,~12.9)\\
        4.244 & 538.1 & 0 &$0.79\pm0.36$  & $0.0^{+0.7}_{-0.0}$&$ (0.0    ,   2.5 )$ &0.422  &0.77 &1.056 &$0.0^{+0.4}_{-0.0} $ &(0.0,~1.4) \\
        4.258 & 828.4 & 6 &$0.88\pm0.46$  & $5.1^{+3.3}_{-2.1}$&$ (1.6    ,  10.9 )$ &0.412  &0.78 &1.054 &$1.8^{+1.2}_{-0.8} $ &(0.6,~3.9) \\
        4.267 & 531.1 & 7 &$1.41\pm0.45$  & $5.6^{+3.3}_{-2.8}$&$ (2.1    ,  11.8 )$ &0.399  &0.79 &1.053 &$3.2^{+1.9}_{-1.6} $ &(1.2,~6.7) \\
        4.278 & 175.7 & 2 &$1.06\pm0.60$  & $0.9^{+2.3}_{-0.8}$&$ ( 0.0   ,    8.0)$ &0.384  &0.82 &1.053 &$1.5^{+3.9}_{-1.4} $ &(0.0,~13.7)\\
        4.308 & 45.1  & 0 &$0.04\pm0.34$  & $0.0^{+1.3}_{-0.0}$&$ ( 0.0   ,    2.4)$ &0.351  &0.94 &1.052 &$0.0^{+8.3}_{-0.0} $ &(0.0,~15.3)\\
        4.358 & 543.9 & 3 &$1.49\pm0.80$  & $1.5^{+2.3}_{-1.2}$&$ (0.0    ,   9.2 )$ &0.281  &1.18 &1.051 &$0.8^{+1.2}_{-0.6} $ &(0.0,~4.9) \\
        4.387 & 55.6  & 0 &$0.00\pm0.31$  & $0.0^{+1.3}_{-0.0}$&$ ( 0.0   ,    2.5)$ &0.252  &1.32 &1.051 &$0.0^{+6.7}_{-0.0} $ &(0.0,~12.9)\\
        4.416 & 1043.9& 8 &$3.03\pm1.02$  & $5.0^{+3.3}_{-2.7}$&$ (1.4    ,  12.4 )$ &0.223  &1.46 &1.052 &$1.4^{+0.9}_{-0.8} $ &(0.4,~3.5) \\
        4.467 & 111.1 & 4 &$0.44\pm0.42$  & $3.6^{+2.7}_{-1.7}$&$ ( 1.2   ,    8.4)$ &0.194  &1.72 &1.055 &$9.2^{+6.9}_{-4.4} $ &(3.1,~21.5)\\
        4.527 & 112.1 & 0 &$0.00\pm0.23$  & $0.0^{+1.3}_{-0.0}$&$ (0.0    ,   2.4 )$ &0.166  &2.02 &1.054 &$0.0^{+3.3}_{-0.0} $ &(0.0,~6.1) \\
        4.575 & 48.9  & 0 &$0.16\pm0.22$  & $0.0^{+1.1}_{-0.0}$&$ ( 0.0   ,    2.4)$ &0.151  &2.25 &1.054 &$0.0^{+6.2}_{-0.0} $ &(0.0,~13.6)\\
        4.600 & 586.9 & 2 &$0.62\pm0.44$  & $1.4^{+2.2}_{-1.0}$&$ (0.0    ,   6.2 )$ &0.143  &2.38 &1.055 &$0.7^{+1.0}_{-0.5} $ &(0.0,~2.9) \\ \hline
        Sum &  & 34       &$10.77\pm1.85$ &&$P$-value& $4.6\times10^{-7}$& &  &Statistical significance &5$\sigma$\\\hline
\end{tabular}
\caption{The cross sections $\sigma^{\rm B}$ and upper limits on $\sigma^{\rm B}$ at the 90\% C.L.\ with the POLE ($\sigma^{\rm B}_{\rm POLE}$) method
for $\EE\to \eta\psip$ at different energy points, together with integrated luminosity $\mathcal{L}_{\rm int}$, numbers of observed events $n^{\rm obs}$,  upper limits
at the 90\% C.L.\ for the number of signal events $n^{\rm obs}_{\rm POLE}$, background events $n^{\rm b}$, and signal events $n^{\rm sig}$, product of detection efficiencies and branching fractions $\Sigma=\mathcal{B}_1\mathcal{B}_2 (\epsilon_e\mathcal{B}_e+\epsilon_\mu\mathcal{B}_\mu)$, ISR correction factor $(1+\delta)$, vacuum polarization factor $|1-\Pi|^{-2}$, the $P$-value, and the statistical significance. The uncertainties of $n^{\rm sig}$ and $\sigma^{\rm B}$ are statistical only. }\label{table:Preliminary results}
\end{sidewaystable*}
Figure~\ref{fig:cross section} shows the measured Born cross sections for
$\EE\to\eta\psip$  as a function of the collision energy.

The $P$-value is obtained by calculating the probability of the expected number of background events to fluctuate to the number of observed events or more in the signal region assuming a Poisson distribution.  The total number of observed events and total expected number of background events are 34 and $10.77\pm1.85$, respectively, in the sum of the 14 data samples at different c.m.\ energies.  Considering the uncertainty of the number of background events,  the $P$-value and the corresponding statistical significance of $\ee\to \eta\psip$ signals from  the $5.25~{\rm fb}^{-1}$ BESIII data  are  $4.6\times10^{-7}$ and   $5\sigma$, respectively,  which are listed in table~\ref{table:Preliminary results}.

\section{Systematic uncertainties}
\label{sec:uncertainty}
The systematic uncertainties in the cross section measurement mainly come from the luminosity, tracking efficiency, photon detection efficiency, kinematic fit, ISR correction factor, mass windows of $\jpsi$, $\eta$, and $\psip$, background estimation, and the branching fractions of intermediate particle decays. The uncertainty from the vacuum polarization is negligible.

\begin{itemize}
\item \emph{Luminosity.} The integrated luminosity was measured using Bhabha scattering events with an uncertainty of $1.0\%$~\cite{luminosity-measurement, luminosity-measurement-2},  which is taken as the systematic uncertainty.

\item \emph{Tracking.} The uncertainty of the tracking efficiency is 1.0\% per track, which is taken from ref.~\cite{intro-bes3-etapjpsi}. 

\item \emph{Photon.} The uncertainty from photon reconstruction is $1.0\%$ per photon, which is determined from the study of the process $\jpsi\to\rho^{0}\piz$, $\rho^{0}\to\pp$, $\piz\to\gamma\gamma$~\cite{photoneffi}.

\item \emph{Kinematic fit.} The systematic uncertainty from the kinematic fit is estimated by correcting the helix parameters of charged tracks according to the method described in ref.~\cite{bes3-kinematicfit-eff}. The MC sample with the track helix parameter correction applied is taken as the nominal one. The difference between detection efficiencies
obtained from MC samples with and without correction is taken as the uncertainty.

\item \emph{ISR correction factor.} Due to insufficient information from previous experiments, we obtain the ISR correction factor according to the decay of the $Y(4260)$ resonant structure in this work. Changing the Breit-Wigner (BW) function for the $Y(4260)$ to that for the $\psi(4415)$, the difference between these two assumptions is taken as the systematic uncertainty.

\item \emph{Branching fraction.} The uncertainties in the branching fractions of $\eta\to \gamma\gamma$, $\psip\to \ppjpsi$ with $\jpsi \to \LL$ are taken from the PDG~\cite{pdg}.

\item \emph{Mass window.} The mass resolution discrepancy between MC simulation and the data will lead to a bias in the efficiency determination when a mass window requirement is applied to the invariant mass distribution. The process $\EE\to \pip\pim\psip$ with $\psip \to \eta \jpsi$ at $\sqrt{s}=4.416$~GeV is taken as the control sample to estimate the uncertainty due to the $\jpsi$ and $\eta$ mass windows. The discrepancies in efficiency between data and MC samples for $\jpsi$ and $\eta$ mass windows are ($0.80\pm0.12$)\% and ($-0.35\pm2.27$)\%, respectively. The uncertainties of $\jpsi$ and $\eta$ mass windows are quoted as 0.92\% and 2.62\%, respectively. The uncertainty of the $\psip$ mass window is determined to be 2.3\%, using a large data sample observed in $\EE\to \pip\pim\psip$~\cite{cross-section-pppsip}. Finally, the total systematic uncertainty on mass windows is 3.6\% by adding these numbers in quadrature.

\item \emph{Background estimation.} From eq.~(\ref{equ:bk number}), the number of background events is estimated using the measured cross sections. We calculate the uncertainty in the number of background events with the uncertainty of the measured cross sections. The ratio of the uncertainty in the number of background events to the number of signal events is taken as the uncertainty of the background estimation.
\end{itemize}

Table~\ref{table:Unscertainties} summarizes the systematic uncertainties from all the sources. The total systematic uncertainty is obtained by summing the individual uncertainties in quadrature, assuming that all sources are independent.
\begin{sidewaystable*}[htbp]
\centering
\begin{tabular}{|cccccccccc|}
\hline
$\sqrt s$~(GeV) & Luminosity& Tracking& Photon&BR&ISR &Kinematic fit&Background&Mass window&Sum\\\hline
    4.236 & 1.0 & 4.0 &2.0& 1.2 & 10.4 &2.6 &23.6 &3.6 &26.6\\
    4.242 & 1.0 & 4.0 &2.0& 1.2 & 10.9 &2.6 &---&3.6 &12.7\\
    4.244 & 1.0 & 4.0 &2.0& 1.2 & 10.5 &2.8 &---&3.6 &12.4\\
    4.258 & 1.0 & 4.0 &2.0& 1.2 & 7.4  &2.7 &9.0 &3.6 &13.4\\
    4.267 & 1.0 & 4.0 &2.0& 1.2 & 5.3  &3.1 &8.0 &3.6 &11.7\\
    4.278 & 1.0 & 4.0 &2.0& 1.2 & 2.4  &3.1 &66.7 &3.6 &67.0\\
    4.308 & 1.0 & 4.0 &2.0& 1.2 & 5.1  &3.2 &---&3.6 &8.5\\
    4.358 & 1.0 & 4.0 &2.0& 1.2 & 10.6 &3.5 &53.3 &3.6 &54.8\\
    4.387 & 1.0 & 4.0 &2.0& 1.2 & 12.4 &3.5 &---&3.6 &14.2\\
    4.416 & 1.0 & 4.0 &2.0& 1.2 & 11.4 &3.1 &20.4 &3.6 &24.3\\
    4.467 & 1.0 & 4.0 &2.0& 1.2 & 2.0  &3.3 &11.7 &3.6 &13.6\\
    4.527 & 1.0 & 4.0 &2.0& 1.2 & 1.2  &3.1 &---&3.6 &6.8\\
    4.575 & 1.0 & 4.0 &2.0& 1.2 & 4.4  &2.8 &---&3.6 &7.9\\
    4.600 & 1.0 & 4.0 &2.0& 1.2 & 5.0  &2.9 &31.4 &3.6 &32.5\\\hline
    \end{tabular}
\caption{The relative systematic uncertainties from luminosity, tracking, photon, branching fraction (BR), ISR  correction factor, kinematic fit, background estimation, and mass windows~(in units of \%). Ellipses mean that the results are not applicable.}\label{table:Unscertainties}
\end{sidewaystable*}

\section{Summary}
\label{sec:summ}
In summary, using $5.25~{\rm fb}^{-1}$ data collected at c.m.\ energies from 4.236 to $4.600~\gev$, the process $\EE\to \eta\psip$ is observed for the first time
with a $5\sigma$ statistical significance. The energy-dependent cross section has been measured and the results are listed in  table~\ref{table:Preliminary results}.
Because of the limited  statistics, the signals at some energy points are not significant, thus it is impossible to extract the couplings of the $Y$ states to $\eta\psip$ from a
fit to the cross sections of $\EE\to \eta\psip$. Further experimental studies with higher statistics are needed to draw a clear conclusion on the structure in the $\EE\to\eta\psip$ process. BESIII plans to collect additional data samples over a variety of c.m. energies in the future~\cite{bes3-white-paper}. Furthermore, a partial event reconstruction technique with a missing track may improve the detection efficiency of this process. This will allow us to study the structure of the $\eta\psip$ and explore the nature of the vector charmonium(like) states.
\clearpage

\acknowledgments

The BESIII collaboration thanks the staff of BEPCII and the IHEP computing center for their strong support. This work is supported in part by National Key Research and Development Program of China under Contracts Nos. 2020YFA0406300, 2020YFA0406400; National Natural Science Foundation of China (NSFC) under Contracts Nos. 11625523, 11635010, 11735014, 11822506, 11835012, 11935015, 11935016, 11935018, 11961141012; the Chinese Academy of Sciences (CAS) Large-Scale Scientific Facility Program; Joint Large-Scale Scientific Facility Funds of the NSFC and CAS under Contracts Nos. U1732263, U1832207; CAS Key Research Program of Frontier Sciences under Contracts Nos. QYZDJ-SSW-SLH003, QYZDJ-SSW-SLH040; 100 Talents Program of CAS; INPAC and Shanghai Key Laboratory for Particle Physics and Cosmology; ERC under Contract No. 758462; European Union Horizon 2020 research and innovation programme under Contract No. Marie Sklodowska-Curie grant agreement No 894790; German Research Foundation DFG under Contracts Nos. 443159800, Collaborative Research Center CRC 1044, FOR 2359, FOR 2359, GRK 214; Istituto Nazionale di Fisica Nucleare, Italy; Ministry of Development of Turkey under Contract No. DPT2006K-120470; National Science and Technology fund; Olle Engkvist Foundation under Contract No. 200-0605; STFC (United Kingdom); The Knut and Alice Wallenberg Foundation (Sweden) under Contract No. 2016.0157; The Royal Society, UK under Contracts Nos. DH140054, DH160214; The Swedish Research Council; U. S. Department of Energy under Contracts Nos. DE-FG02-05ER41374, DE-SC-0012069.



\end{document}